\documentclass[10pt,conference]{setup/IEEEtran}
\usepackage[utf8]{inputenc}

\usepackage{cite}
\usepackage{amsmath,amssymb,amsfonts,mathtools,bm,etoolbox}
\usepackage{algorithmic}
\usepackage{graphicx}
\usepackage{textcomp} % Copyright and other symbols
\usepackage[nolist]{acronym}
\usepackage{siunitx} % \SI package.
\usepackage{xcolor} % Defining colours
\usepackage{upgreek} %upright lowercase greek symbols e.g. \uplambda which is not included in latex default.
\usepackage{commath} %Provides easier interface for abs value and norm, \dif x (differential operator), \od[]{f}{x} (ordinary derivative), \pd[2]{f}{x} (partial derivative), and mixed derivative \md.

\usepackage{multicol}
\usepackage{subcaption}

\definecolor{orange}{RGB}{255, 165, 0}

\colorlet{darkblue}{blue!50!black}
\colorlet{orange}{red!50!yellow}

\newcommand{\e}[1]{\mathrm{e}^{#1}}

\newcommand{\Int}{\int\limits}

\renewcommand{\vec}[1]{\mathbf{\lowercase{#1}}}
\newcommand{\mat}[1]{\mathbf{\uppercase{#1}}}

\let\oldRe\Re
\renewcommand{\Re}[1]{\oldRe\left(#1\right)}
\newenvironment{units}
{ Where:\\ }
{ \\ }
\newcommand{\punkt}[3]{
\begin{tabular}{ p{10pt} l p{170pt} l}
~~{#1} &~=~&{#2} &{$\left[\text{#3}\right]$}
\end{tabular}
\\
}
\renewcommand{\punkt}[3]{
\begin{tabular}{ p{10pt} l p{202pt}}
~~{#1} &~=~&{#2}
\end{tabular}
\\
}

\DeclarePairedDelimiterXPP\Aver[1]{\mathbb{E}}{[}{]}{}{

#1
}

\DeclareMathOperator*{\argmax}{arg\,max} % thin space, limits underneath in displays

\acrodef{LIS}{Large Intelligent Surface}
\acrodef{UE}{User Equipment}
\acrodef{RMS}{Root Mean Square}
\acrodef{AWGN}{Additive White Gaussian Noise}
\acrodef{SNR}{Signal-to-Noise Ratio}
\acrodef{nCA-MF}{non-Coupling-Aware Matched Filtering}
\acrodef{CA-MF}{Coupling-Aware Matched Filtering}
\acrodef{nCA-ZF}{non-Coupling-Aware Zero Forcing}
\acrodef{CA-ZF}{Coupling-Aware Zero Forcing}
\acrodef{SVD}{Singular Value Decomposition}
\acrodef{EMS}{Electromagnetic Surface}
\acrodef{MF}{Matched Filtering}

\title{A Communication Model for Large Intelligent Surfaces}
\author{\IEEEauthorblockN{Robin~Jess~Williams$^1$, Elisabeth~De~Carvalho$^1$, Thomas L. Marzetta$^2$}
\IEEEauthorblockA{$^1$Department of Electronic Systems, Aalborg University, Denmark \\
$^2$ Tandon School of Engineering, New York University, Brooklyn, NY   \\
Email: $^1$\{rjw, edc\}@es.aau.dk, $^2$tom.marzetta@nyu.edu} \\
}
\begin{document}

\maketitle

\begin{abstract}
The purpose of this paper is to introduce a communication model for Large Intelligent Surfaces (LIS).  A LIS is modeled as a collection of tiny closely spaced antenna elements. Due to the proximity of the elements, mutual coupling arises. 
An optimal transmitter design depends on the mutual coupling matrix. For single user communication, the optimal transmitter uses the inverse of the mutual coupling matrix in a filter matched to the channel vector. We give the expression of the mutual coupling for two types of planar arrays. The conditioning number of the mutual coupling matrix is unbounded as the antenna element density increases, so only the dominant values can be inverted within reasonable computation. The directivity is partial but still significant compared to the conventional gain. When the spacing between elements becomes small (smaller than half a wavelength), the directivity surpasses the conventional directivity equal to the number of antennas, as well as the gain obtained when modeling the surface as continuous. The gain is theoretically unbounded as the element density increases for a constant aperture. \end{abstract}

\section{Introduction}\label{sec:introduction}
The introduction of Massive multiple-input multiple-output (MIMO) systems \cite{marzetta2010} has defined a new generation of base stations that are equipped with a very large number of antennas. Evolving from massive MIMO, new visions have emerged where the base stations are equipped with an even larger number of antennas. In \cite{carvalhoHeath}, the focus is put on increasing the dimension of the antenna arrays from moderately large in conventional massive MIMO systems to extremely large arrays, possibly tens of meters. This type of system is called extra-large scale MIMO (XL-MIMO) or an Extremely Large Aperture Array (ELAA) system. The idea relies on the possibility of creating yet again a new generation of base stations that are easy to deploy and are easily integrable in the surroundings. 
Recent developments in array design point towards the feasibility of this vision with low cost, low weight and ultra-thin arrays \cite{abs-1901-01458,prather}.

We designate those ultra-thin electromagnetic sheets as \acp{LIS} as introduced in \cite{lund}, i.e. a large electromagnetic surface that is active and able to transmit and receive electromagnetic waves.
Very few papers exist on the research topic of active \acp{LIS} and little is known about their performance in terms of communications. In particular, it is essential to provide realistic models that are close to reality and can support transceiver design as well as performance assessment. This paper aims at providing such a model. In \cite{lund}, an \ac{LIS} is modelled as a continuous surface of infinitesimal plane antennas and does not capture one essential characteristic about the propagation of current within the surface. 

We model a surface as a collection of tiny closely spaced antenna elements. Because of the proximity of the antennas elements, mutual coupling arises \cite{whatIsNext}. This feature is essential in \acp{LIS} and is the main focus of this paper. The impact of mutual coupling is not only about an appropriate modelling of the \ac{LIS}. It changes the optimization of transceivers and opens the possibility of major performance improvements. 
We provide the expression of the radiated power as a function of the impedance matrix. This step is important as it gives the transmit power constraint necessary in transceiver design. Furthermore, in a single user communication scenario, we provide the expression of the optimal transmitter which is the whitened matched filter. 
We investigate two models: one that is based on a discrete model with isotropic antennas and one that is a collection of closely spaced planar antenna elements. For both models, we derive the expression of the impedance matrix. 
For planar arrays, we investigate the performance using a fixed aperture as the array is populated by a higher and higher number of antennas. As the antenna element density increases, the directivity grows unbounded. In practice, however, performance is limited but remains very promising. The optimal processing involves the inverse of the impedance matrix. The impedance matrix exhibits very small singular values and cannot be inverted within reasonable computational precision. However, by inverting the dominant eigenvalues, a significant portion of the directivity can be captured.

Achieving those superior gains comes with long-standing challenges. These challenges include high ohmic losses due to high currents \cite{Schelkunoff}, the need for very precise adjustment of the excitation currents, and scan blindness. Solutions have been proposed for some of these challenges \cite{balanis}. 
One important message~\cite{whatIsNext} is that those challenges should not stand in the way of super-directivity. Those gains naturally pertain to the \ac{LIS} technology. They will be uncovered by upcoming engineering progress. 

\textit{Paper overview:} The rest of the paper is organized as follows: Section \ref{sec:systemModel} presents the scenario model of the \ac{LIS}. Subsections \ref{subSec:rxPower}, \ref{subSec:CFIso}, and \ref{subSec:CFFlat} details the signal received at the \ac{UE} and impedance functions for two different models of the \ac{LIS}.  Section \ref{sec:singleUser} applies the derived channel to a single-user system and investigates the effect of the inter-element coupling. Section \ref{sec:simResults} provides simulation results of the single-user system. Section \ref{conclusion} concludes this paper.

\textit{Notation:} $j$ is the imaginary unit, $\Vert\cdot\Vert_2$ is the euclidean norm, $\vert\cdot\vert$ is the absolute value, $\cdot^*$ is the complex conjugate, and $\cdot^T$ and $\cdot^H$ are the transpose and hermitian transpose respectively. $\cdot^{-1}$ and $\cdot^\dagger$ is the matrix inverse and matrix pseudo-inverse respectively. Vectors are denoted by bold lowercase symbols. Matrices are denoted by bold uppercase symbols.

\section{System model: LIS and mutual coupling}\label{sec:systemModel}
\begin{figure}[h]
    \centering
    \includegraphics[scale=1]{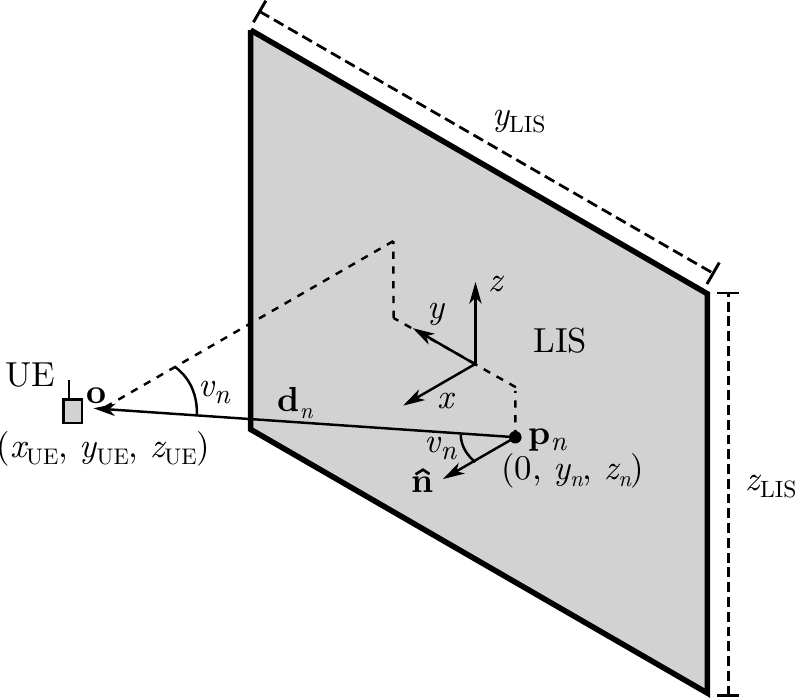}
    \caption{Illustration of the scenario with a \ac{LIS} of height $z_\text{LIS}$ and width $y_\text{LIS}$. The angle of arrival between a \ac{UE} at point $\vec{o}$ and an antenna element at point $\vec{p}_n$ is denoted $v_{n}$. $\vec{d}_n$ is the distance vector between the \ac{UE} and the $n$'th element.}
    \label{fig:scenario}
\end{figure}

We consider a \ac{LIS} system as depicted in Fig. \ref{fig:scenario}. 
We investigate a simple scenario where the \ac{LIS} transmits to a single user in line-of-sight propagation. The reason for choosing a simple scenario is that we want to highlight the super-directivity properties of a \ac{LIS} in a simple system first before introducing a more complex system with multiple users and a more advanced channel model. 
The  surface is placed on the wall of a large indoor venue. 
The \ac{LIS} consists of $N$ identical and equally spaced antenna elements within an area of height $z_\text{LIS}$ and width $y_\text{LIS}$ and is centred at the origin. 
The \ac{LIS} transmits signals to a single \ac{UE} equipped with a single isotropic antenna.

The distance between the \ac{UE} at position $\vec{o} = \begin{bmatrix} x_\text{UE} & y_\text{UE} & z_\text{UE}\end{bmatrix}^T$  and the $n$'th element of the \ac{LIS} at position $\vec{p}_n = \begin{bmatrix} 0 & y_n & z_n\end{bmatrix}^T$ is given by Eq. \eqref{eq:dist} and the angle of departure is given by Eq. \eqref{eq:AoA}.
\begin{equation}
    d_{n} = \left\Vert \vec{d}_n \right\Vert_2  = \left\Vert \vec{o} - \vec{p}_n \right\Vert_2 \label{eq:dist}\\
\end{equation}
\begin{equation}
    v_{n} = \arccos\left(\frac{x_\text{UE}}{d_{n}}\right) \label{eq:AoA}
\end{equation}
When the inter-element spacing in the \ac{LIS} is low, mutual coupling causes current from one antenna to induce current in the adjacent antennas. This effect is important to account for in assessing the performance of a communication system with respect to the radiated power. In the next section, the power received at the user is presented, and the impedance matrices are introduced.

\subsection{Received power and impedance matrices} \label{subSec:rxPower}
For this model, all fields and antennas are assumed to be polarized identically. Due to this assumption, the vector fields are represented as scalar fields.
The total field strength at the \ac{UE} is given as the sum of contributions from all $N$ transmitting antenna elements as stated by Eq. \eqref{eq:sumSignal}. \cite{mailloux}
\begin{align}
 E(\vec{o}) =  \sum_{n = 1}^N  E_n(\vec{o}, \vec{p}_n) \label{eq:sumSignal}
\end{align}
The contribution from the $n$'th antenna element is given by Eq. \eqref{eq:singleSignal}. All antenna elements of the \ac{LIS} are assumed identical. For this reason, $G$ and $\beta$ do not carry the $n$ subscript.
\begin{align}
    E_n(\vec{o}, \vec{p}_n) =  \sqrt{\beta} \sqrt{ \frac{G(\theta_n, \phi_n) \eta}{4 \pi}} I_n \frac{\e{-j k d_{n}}}{d_{n}} \label{eq:singleSignal}
\end{align}
Combining Eqs. \eqref{eq:sumSignal} and \eqref{eq:singleSignal}, the total field strength can be expressed in matrix notation as Eq. \eqref{eq:sumSignalMatrix}.
\begin{align}
 E(\vec{o}) =  \sum_{n = 1}^N  E_n(\vec{o}, \vec{p}_n) = \sqrt{\eta}\sqrt{\frac{4\pi\beta}{\lambda^2}} \vec{i}^H \vec{h} \label{eq:sumSignalMatrix}
\end{align}
\begin{units}
    \punkt{$\vec{I}$}      {$N$x1 excitation vector with entries $\vec{i}_n=I_n$}  {\si{\ampere}}
    \punkt{$\vec{h}$}      {$N$x1 channel vector with entries $\vec{h}_{n}=\sqrt{G\left(\theta_n,\phi_n\right)} \frac{\lambda}{4\pi d_{n}}\e{-j k d_{n}}$}  {1}
    \punkt{$I_n$}     {complex excitation current for antenna $n$}  {\si{\ampere}}
    \punkt{$\beta$}     {proportionally factor accounting for antenna characteristics}  {\si{\ampere}}
    \punkt{$\eta$}      {intrinsic impedance of vacuum}  {1}
    \punkt{$k$}{wavenumber given as $\frac{2\pi}{\lambda}$}{}
    \punkt{$\lambda$}{wavelength in vacuum}{}
    \punkt{$G$}{ antenna gain of the $n$'th antenna}{}
    \punkt{$\theta_n$}{polar angle towards the \ac{UE} from the $n$'th antenna}{}
        \punkt{$\phi_n$}{azimuth angle towards the \ac{UE} from the $n$'th antenna}{}
\end{units}
The surface power density at the \ac{UE} is given by Eq. \eqref{eq:singlePowerD}.%(angular dependency of gains omitted to shorten notation).
\begin{align}
    &W(\vec{o}) = \frac{ E(\vec{o}) E(\vec{o})^* }{\eta} = \frac{4\pi\beta}{\lambda^2} \vec{i}^H\vec{h}\vec{h}^H\vec{i} =  \frac{4\pi\beta}{\lambda^2} \left\vert \vec{i}^H\vec{h} \right\vert^2 \label{eq:singlePowerD} \\
        &\phantom{~~} = \frac{\beta}{4\pi} \sum_{n=1}^N \sum_{m=1}^N \sqrt{G(\theta_n,\phi_n) G(\theta_m,\phi_m)} I_n I_m^* \frac{\e{-jk\left(d_{n}-d_{m}\right)}}{d_{n}d_{m}}\notag
\end{align}
The received power at the \ac{UE} is equal the product of the surface power density and the effective aperture of the receiving antenna as stated in Eq. \eqref{eq:singlePower}, assuming that the field is constant across the dimensions of the receiver. \cite{parsons}
\begin{align}
    P = W(\vec{o}) A = \frac{4\pi\beta}{\lambda^2} \left\vert \vec{i}^H\vec{h} \right\vert^2 \frac{\lambda^2}{4\pi}  = \beta \left\vert \vec{i}^H\vec{h} \right\vert^2\label{eq:singlePower}
\end{align}
\begin{units}
    \punkt{$A$}      {the effective aperture of the receiving antenna}  {\si{\ampere}}
\end{units}
To express the received power relative to the transmitted power, the total radiated power is derived. Since electric fields are solenoidal \cite{griffiths}, the total radiated power can be calculated by integration of the power density over any closed surface which encloses the \ac{LIS}. In Eq. \eqref{eq:totalRadPower} the surface is chosen as a sphere with radius $r$.
\begin{align}
    P_\text{rad} = \Int_{0}^{2\pi}\Int_{0}^{\pi} W(r\hat{\vec{r}}) r^2 \sin\left(\theta\right) \dif \theta \dif \phi \label{eq:totalRadPower}
\end{align}
\begin{units}
    \punkt{$\hat{\vec{r}}$}      {$\begin{bmatrix} \cos\left(\phi\right)\sin\left(\theta\right)& \sin\left(\phi\right)\sin\left(\theta\right) & \cos\left(\theta\right)\end{bmatrix}^T$}  {\si{\ampere}}
\end{units}
As the radiated power is independent of the integration surface, the radius of the sphere is chosen so large that the angle between the \ac{LIS} normal $\hat{\vec{n}}$ and the distance vector $\vec{d}_n$ for all antennas on the \ac{LIS} to any point on the integration sphere is approximately equal. (see Fig. \ref{fig:scenario}) Under these conditions, the power density can be approximated as Eq. \eqref{eq:singlePowerDFF} \cite{mailloux}.
\begin{align}
    W_{FF}(\theta,\phi,r) = \frac{\beta}{4\pi} \sum_{n=1}^N \sum_{m=1}^N G\left(\theta,\phi\right) I_n I_m^* \frac{\e{-ik \hat{\vec{r}}^H\left(\vec{p}_{n}-\vec{p}_{m}\right)}}{r^2} \label{eq:singlePowerDFF}
\end{align}
The total radiated power can then be expressed as Eq. \eqref{eq:totalRadPowerFF}.
\begin{equation}
    P_\text{rad} = \beta \sum_{n=1}^N \sum_{m=1}^N I_n I_m^* z(\vec{p}_n,\vec{p}_m) = \beta \vec{i}^H \mat{Z} \vec{i} \label{eq:totalRadPowerFF}
\end{equation}
\begin{equation}
    z(\vec{p}_n, \vec{p}_m) = \Int_{0}^{2\pi}\Int_{0}^{\pi} \frac{G\left(\theta,\phi\right)}{4\pi} \e{-ik \hat{\vec{r}}^H\left(\vec{p}_{n}-\vec{p}_{m}\right)} \sin\left(\theta\right) \dif \theta \dif \phi \notag
\end{equation}
\begin{units}
\punkt{$z$}{impedance function}{}
    \punkt{$\mat{Z}$}      {$N$x$N$ impedance matrix with entries $\mat{Z}_{n,m}=z\left(\vec{p}_n, \vec{p}_m\right)$}  {1}
\end{units}
Combining Eqs. \eqref{eq:totalRadPowerFF} and \eqref{eq:singlePower} yields the power received at the \ac{UE} as Eq. \eqref{eq:singlePowerRx}.
\begin{align}
    P = P_\text{rad} \frac{\left\vert \vec{i}^H\vec{h} \right\vert^2}{\vec{i}^H\mat{Z}\vec{i}} \label{eq:singlePowerRx}
\end{align}

In the following two sections, the coupling function is derived for isotropic and planar antenna elements respectively. 

\subsection{Impedance function for isotropic elements}\label{subSec:CFIso}
\begin{figure}
    \centering
    \includegraphics[scale=1]{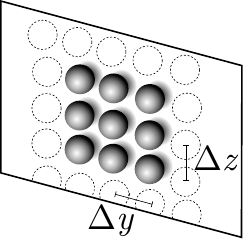}
    \caption{Illustration of \ac{LIS} made from isotropic antenna elements pictured as spheres. The antenna elements are spaced $\Delta y$ along the $y$-axis and $\Delta z$ along the $z$-axis.}
    \label{fig:LIS_Isotropic}
\end{figure}

The gain pattern of an isotropic antenna is given as $G(\theta,\phi) = 1$. Fig. \ref{fig:LIS_Isotropic} shows an illustration of the \ac{LIS} made from isotropic antennas.

With unit gain, the entries of the channel vector is given by $\vec{H}_{n} = \frac{\lambda}{4\pi d_{n}}\e{-i k d_{n}}$ and the impedance function is given by Eq. \eqref{eq:isotropicCoupling}.
\begin{align}
    &z_\text{iso}(\vec{p}_n,\vec{p}_m) = \Int_{0}^{2\pi}\Int_{0}^{\pi} \frac{\e{-ik \hat{\vec{r}}^H\left(\vec{p}_{n}-\vec{p}_{m}\right)}}{4\pi}  \sin\left(\theta\right) \dif \theta \dif \phi \notag \\
    &\phantom{z_{iso}(\vec{p}_n,\vec{p}_m)} = \frac{\sin\left(k\left\Vert \vec{p}_n - \vec{p}_m\right\Vert_2\right)}{k\left\Vert \vec{p}_n - \vec{p}_m\right\Vert_2}\label{eq:isotropicCoupling}
\end{align}

\subsection{Impedance function for planar elements}\label{subSec:CFFlat}
\begin{figure}
    \centering
    \includegraphics[scale=1]{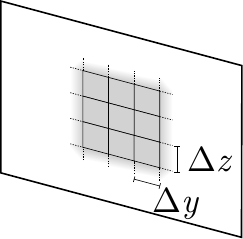}
    \caption{Illustration of \ac{LIS} made from closely spaced planar antenna elements. Each element has an aperture of $A = \Delta y \Delta z$. The centre-to-centre spacing along the $z$-axis and $y$-axis is $\Delta z$ and $\Delta y$ respectively.}
    \label{fig:LIS_2D}
\end{figure}

Fig. \ref{fig:LIS_2D} shows an illustration of the \ac{LIS} made from planar antenna elements. 
The gain pattern of a planar element is given as Eq. \eqref{eq:effectiveArea}.
\begin{align}
    G(\theta,\phi) = \frac{4\pi A_p(\theta,\phi)}{\lambda^2}\label{eq:effectiveArea}
\end{align}
The perceived aperture $A_p$ of an element is equal the actual aperture $A$ projected onto the plane perpendicular to the direction of departure. The effective area is therefore dependent on the angle between the normal vector of the antenna element and the direction of departure. Assuming the size of an element is sufficiently small, so the variation in the angle of departure across the element is negligible, the perceived area is given by Eq. \eqref{eq:projectedArea}.
\begin{align}
    A_p(\theta, \phi) = A \cos\left(v\right) = A \left\vert\sin\left(\theta\right)\cos\left(\phi\right)\right\vert \label{eq:projectedArea}
\end{align}
Using Eq. \eqref{eq:projectedArea}, the impedance function is given as Eq. \eqref{eq:flatCoupling}.
\begin{align}
    &z_\text{pla}(\vec{p}_n,\vec{p}_m) = \Int_{0}^{2\pi}\Int_{0}^{\pi} \frac{A_p(\theta,\phi)}{\lambda^2}  \e{-ik \hat{\vec{r}}^H\left(\vec{p}_{n}-\vec{p}_{m}\right)}  \sin\left(\theta\right) \dif \theta \dif \phi \notag \\
    &\phantom{z_\text{flat}(\vec{p}_n,\vec{p}_m)} = \frac{4\pi A}{\lambda^2} \frac{J_1(k \left\Vert \vec{d}_n - \vec{d}_m \right\Vert_2)}{k \left\Vert \vec{d}_n - \vec{d}_m \right\Vert_2} \label{eq:flatCoupling}
\end{align}
Since both the channel and impedance matrix coefficients contain a constant term which cancels when determining the received power, the $\frac{4\pi A}{\lambda^2}$ term is removed from both expressions. The impedance matrix entries are then given by Eq. \eqref{eq:flatCoupling2}. For the channel vector far-field cannot be assumed, so by using the expression for the angle of departure in Eq. \eqref{eq:AoA}, the channel matrix entries are given by Eq. \eqref{eq:flatChannel}.
\begin{equation}
    \mat{Z}_{n,m} = \frac{J_1(k \left\Vert \vec{d}_n - \vec{d}_m \right\Vert_2)}{k \left\Vert \vec{d}_n - \vec{d}_m \right\Vert_2} \label{eq:flatCoupling2}
    \end{equation}
    \begin{equation}
    \vec{h}_{n} = \sqrt{\frac{x_\text{UE}}{d_{n}}} \frac{\lambda}{4\pi d_{n}}\e{-i k d_{n}} \label{eq:flatChannel}
\end{equation}
\begin{units}
    \punkt{$J_1$} {first order bessel function of the first kind}   {}
\end{units}

The following section applies the derived impedance matrices to evaluate the effect of the inter-element coupling for a single user communication system.

\section{Single-user system}\label{sec:singleUser}
The signal received at the \ac{UE} for a free-space \ac{AWGN} channel is given by Eq. \eqref{eq:singleUserChannel}.
\begin{align}
    y =  x \hat{\vec{i}}^H \vec{h}  + n \label{eq:singleUserChannel}
\end{align}
\begin{units}
    \punkt{$x$} {transmitted symbol to the \ac{UE} modeled as a random variable with mean $\mu_x=0$ and variance $\sigma_x^2=P_{tx}$ }   {}
    \punkt{$n$} {realization of complex white Gaussian noise variable $N$ with mean $\mu_n=0$ and variance $\sigma_n^2$}    {}
    \punkt{$\hat{\vec{i}}$} {$N$x1 power-normalized precoding vector}    {}
    \punkt{$\vec{H}$}     {$N$x1 free-space channel vector}    {}
\end{units}
The power-normalized precoding vector is given as Eq. \eqref{eq:powerNorm}
\begin{align}
    \hat{\vec{i}} = \frac{\vec{i}}{\sqrt{\vec{i}^H \mat{Z} \vec{i}}} \label{eq:powerNorm}
\end{align}
\begin{units}
    \punkt{$\vec{i}$} {$N$x1 precoding vector}   {}
    \punkt{$\mat{z}$} {$N$x$N$ impedance matrix}   {}
\end{units}
The \ac{SNR} is given by Eq. \eqref{eq:singleUserSNR}.
\begin{align}
        \text{SNR} =   \frac{P_{tx}}{\sigma_n^2} \frac{\vec{i}^H\vec{h}\vec{h}^H\vec{i}}{\vec{i}^H \mat{Z} \vec{i}}  =  \frac{P_{tx}}{\sigma_n^2} \frac{\left\vert \vec{i}^H\vec{h} \right\vert^2}{\vec{i}^H\mat{Z}\vec{i}}
    \label{eq:singleUserSNR}
\end{align}

The power received from the \ac{LIS} relative to a case with a single isotropic transmitter is given by Eq. \eqref{eq:arrayGain} and is used as the performance metric in the coming sections. For a distant \ac{UE}, this quantity is equal to the directivity of the \ac{LIS} and is therefore referred to as the directivity for the remainder of this paper.
\begin{align}
    \mathit{D}\left(\vec{o}\right) = \frac{\vec{i}^H\vec{h}\vec{h}^H\vec{i}}{\vec{i}^H\mat{z}\vec{i}} \left(\frac{4\pi \left\Vert \vec{o}\right\Vert_2}{\lambda}\right)^2 \label{eq:arrayGain}
\end{align}
The power normalized precoding vector is chosen based on two coding schemes:
\begin{enumerate}
    \item \ac{nCA-MF}
    \item \ac{CA-MF}
\end{enumerate}

\subsection{\ac{nCA-MF}}

For \ac{nCA-MF}, the precoding vector is chosen as the hermitian transpose of the channel. The \ac{nCA-MF} precoding vector is given by Eq. \eqref{eq:singleUsernCAMF}.
\begin{align}
    \vec{i}_\text{nCA-MF} = \vec{H} \label{eq:singleUsernCAMF}
\end{align}
The \ac{SNR} is then given by Eq. \eqref{eq:singleUserSNRnCAMF}.
\begin{align}
    \text{SNR}_\text{nCA-MF} = \frac{\vec{H}^H\vec{H}\vec{H}^H\vec{H}}{\vec{H}^H \mat{Z} \vec{H}} \frac{P_{tx}}{\sigma_n^2}  \label{eq:singleUserSNRnCAMF}
\end{align}

\subsection{\ac{CA-MF}}

For \ac{CA-MF}, the precoding vector is chosen to optimize the \ac{SNR} as stated in Eq. \eqref{eq:singleUserCAMF} \cite{Richards2014}.
\begin{align}
    \vec{i}_\text{CA-MF}=\argmax_\vec{i}  \frac{\vec{i}^H\vec{h}\vec{h}^H\vec{i}}{\vec{i}^H \mat{Z} \vec{i}} = \mat{Z}^{-1} \vec{h}    \label{eq:singleUserCAMF}
\end{align}
The \ac{SNR} is then given by Eq. \eqref{eq:singleUserSNRCAMF}.
\begin{align}
    \text{SNR}_\text{CA-MF} = \vec{H}^H \mat{Z}^{-1} \vec{H} \frac{P_{tx}}{\sigma_n^2}  \label{eq:singleUserSNRCAMF}
\end{align}
In the case where the coupling is non-existent, where $\mat{Z}$ equals the identity matrix, Eqs. \eqref{eq:singleUserSNRnCAMF} and \eqref{eq:singleUserSNRCAMF} are equal.

Calculating the \ac{CA-MF} precoding vector in Eq. \eqref{eq:singleUserCAMF} requires inversion of the matrix $\mat{Z}$. As the inter-element spacing is reduced below $\frac{\lambda}{2}$ the conditioning number of $\mat{Z}$ increases exponentially which limits the computability of the matrix inverse. 

\begin{figure}[h]
    \centering
    \includegraphics{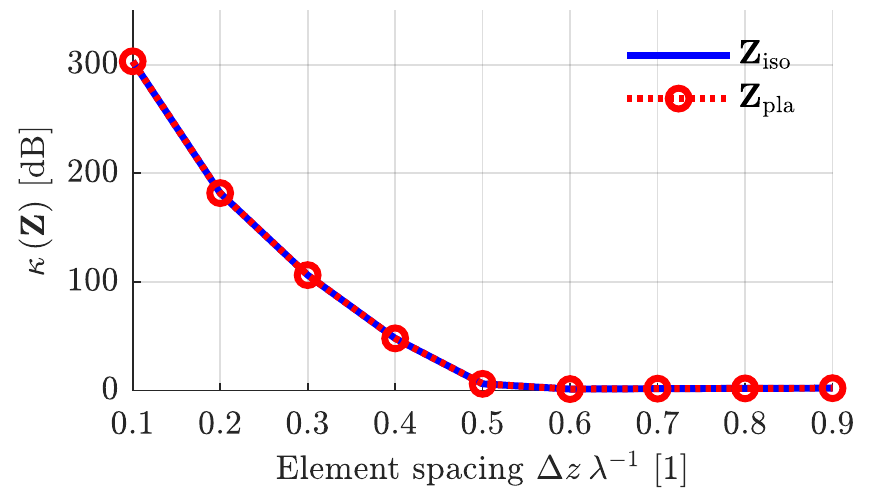}
    \caption{Conditioning number of the coupling matrices for a 20 element linear array. The conditioning number increases significantly as the spacing $\Delta z$ goes below $\frac{\lambda}{2}$.}
    \label{fig:condNum}
\end{figure}

\begin{figure}[h]
    \centering
    \includegraphics[scale=1]{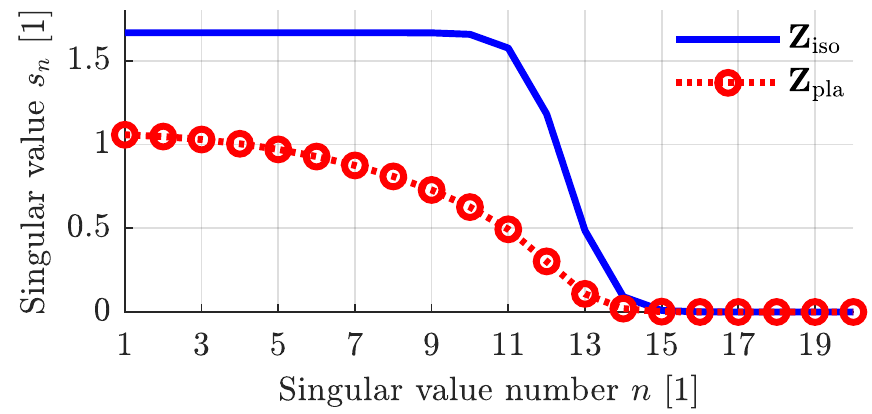}
    \caption{Profile of the singular values of the coupling matrices for a 20 element linear array with element spacing $\Delta z = 0.3\lambda$.}
    \label{fig:singularValues}
\end{figure}

This is illustrated through an example with a linear array of 20 antenna elements placed on the $z$-axis, with a spacing of $\Delta z$. Fig. \ref{fig:condNum} shows the conditioning number of the impedance matrix given by Eq. \eqref{eq:condNum} as the spacing $\Delta z$ is varied. Fig. \ref{fig:singularValues} shows a profile of the singular values for a spacing of $\Delta z = 0.3\lambda$.
\begin{equation}
    \kappa\left(\mat{Z}\right) = \frac{s_\text{max}}{s_\text{min}} \label{eq:condNum}
\end{equation}
\begin{units}
    \punkt{$s_{\text{max}}$} {maximum singular value of $\mat{Z}$}   {}
    \punkt{$s_{\text{min}}$} {minimum singular value of $\mat{Z}$}   {}
\end{units}
An alternative is to use the truncated SVD of $\mat{Z}$ - Limiting the conditioning number, but also reducing the directivity. The \ac{SVD} of the positive definite impedance matrix can be expressed as Eq. \eqref{eq:precodeSVD}.
\begin{equation}
   \mat{Z}=  \mat{U} \mat{S} \mat{U}^H  = \sum_{n=1}^N s_n \vec{u}_n \vec{u}_n^H \label{eq:precodeSVD}
\end{equation}
\begin{units}
    \punkt{$\mat{S}$} {$N$x$N$ diagonal matrix containing singular values $s_1$ to $s_N$ sorted in a descending fashion}   {}
    \punkt{$\mat{U}$} {$N$x$N$ unitary matrix containing the singular vectors $\vec{u}_n$ corresponding to the singular values $s_n$}    {}
\end{units}

\vspace{20pt}
\vspace{20pt}
\begin{figure}[!h]
    \centering
    \includegraphics[scale=1]{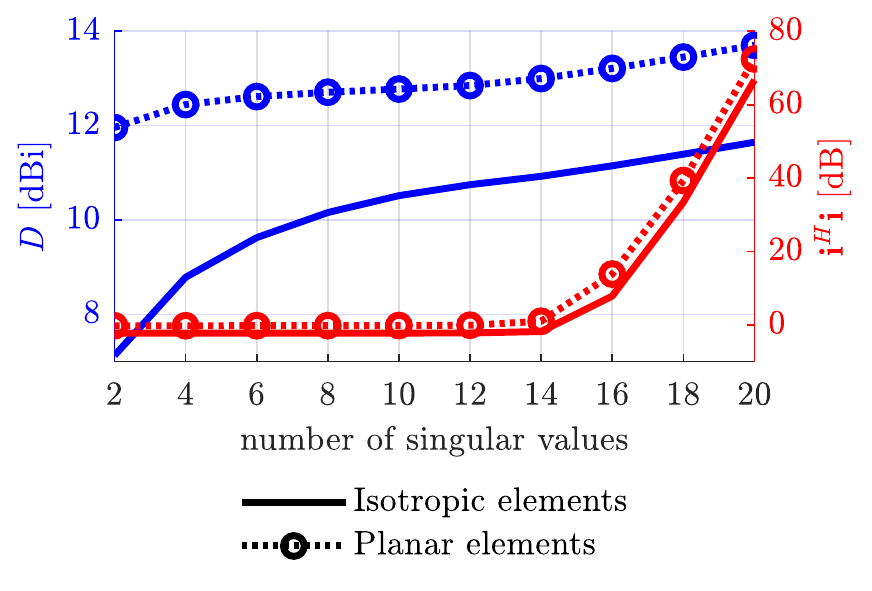}
    \caption{Directivity and current dependency on the number of inverted singular values for a 20 element linear array and a \ac{UE} in the broadside direction.}
    \label{fig:DirectivityCurrents}
\end{figure}

The inverted truncated SVD of $\mat{Z}$ can be calculated by only inverting the singular values within a set threshold, as stated by Eq. \eqref{eq:precodeSVD2}.
\begin{equation}
   \mat{Z}^\dagger = \mat{U} \mat{S}^\dagger \mat{U}^H  = \sum_{n=1}^{N-M} s_n^{-1} \vec{u}_n \vec{u}_n^H \label{eq:precodeSVD2}
\end{equation}
\begin{units}
    \punkt{$M$} {number of singular values below a set threshold}   {}
\end{units}
Using the truncated inverse, the precoding vector and \ac{SNR} is given by \eqref{eq:precodePinv} and \eqref{eq:SNRPinv} respectively.
\begin{equation}
    \vec{v}_\text{CA-pMF}= \mat{Z}^{\dagger} \vec{H}    \label{eq:precodePinv}
\end{equation}
\begin{equation}
    \text{SNR}_\text{CA-pMF} = \vec{H}^H \mat{Z}^{\dagger} \vec{H} \frac{P_{tx}}{\sigma_n^2}  \label{eq:SNRPinv}
\end{equation}
Fig. \ref{fig:DirectivityCurrents} shows the directivity and the power of the excitation currents $\vec{i}^H\vec{i}$ with respect to the number of singular values used. The last few singular values significantly increase the excitation currents. To excite the modes corresponding to the numerically lowest singular values, and realise the maximum directivity, the antenna elements must carry a very high current which is inconvenient for physical systems with ohmic losses. The excitation currents increase as spacing $\Delta z$ approaches zero. For $\Delta z \geq \frac{\lambda}{2}$, the singular values are well above zero and the excitation currents remain low.

\section{Simulation results} \label{sec:simResults}
The \ac{LIS} has a size of $y_\text{LIS} = z_\text{LIS} = \SI{0.5}{\meter}$. The centre frequency is chosen as $f=\SI{2.6}{\giga\hertz}$. The size of the \ac{LIS} is held constant as the inter-element spacing $\Delta y = \Delta z = \Delta d$ is varied. As the spacing decreases, the \ac{LIS} is populated by a higher and higher number of antennas. 

A \ac{UE} is positioned at a point, $\vec{o} = \begin{bmatrix} 10 & 0 & 0 \end{bmatrix}^T$, 10 meters from the centre of the \ac{LIS}. At a distance of 10 meters, the variation in distance between the \ac{UE} and every individual antenna element of the \ac{LIS} is negligible relative to the wavelength. Because of this, the directivity given by Eq. \eqref{eq:arrayGain} remains constant as the distance increases beyond this point.

To illustrate the effect of the mutual coupling the results are compared to Eq. \eqref{eq:effectiveAreaIntegral} \cite{8319526}, which describes the \acs{MF} directivity of a continuous surface of infinitesimal plane antenna elements where mutual coupling is ignored. 
\begin{align}
    \mathit{D}_\text{NC} = \left(\frac{4\pi \left\Vert \vec{o}\right\Vert_2}{\lambda}\right)^2 \Int_{-y_\text{LIS}}^{y_\text{LIS}} \Int_{-z_\text{LIS}}^{z_\text{LIS}} \frac{x_\text{UE}}{4\pi d_p^3} \dif z \dif y \label{eq:effectiveAreaIntegral}
\end{align}

\begin{figure}
    \centering
    \includegraphics{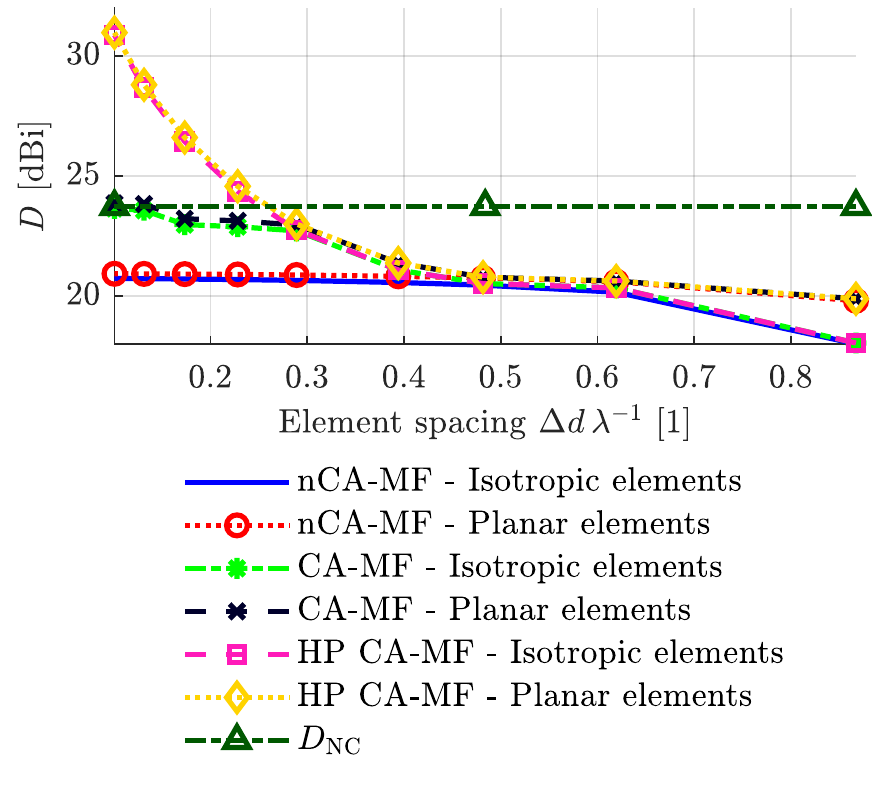}
    \caption{Planar \ac{LIS} directivity in broadside. The physical aperture is constant, so as element spacing decreases, the array is populated by an increasing number of antennas. The performance is shown for limited precision matrix inversion \ac{CA-MF}, and High Precision (HP) CA-MF}
    \label{fig:singleUserBroadside}
\end{figure}

Fig. \ref{fig:singleUserBroadside} shows the directivity for the two coding schemes and antenna element types. 
The \ac{nCA-MF} and \ac{CA-MF} precoding schemes are found to provide almost identical performance for element spacings $\Delta d \geq \frac{\lambda}{2}$, where the inter-element coupling is weak. The planar elements provide superior performance in the broadside, due to the more directional radiation pattern of the individual elements. 

For spacings $\Delta d < \frac{\lambda}{2}$, where the coupling is strong, the isotropic and planar elements provide nearly the same performance. The \ac{nCA-MF} performance is observed to approach a limit as the array is populated by a higher and higher number of antennas.

The directivity for the \ac{CA-MF} scheme grows unbounded given infinite precision - See High Precision (HP) CA-MF line in Fig. \ref{fig:singleUserBroadside}. Limiting the precision by only inverting singular values higher than $s_\text{min} = 10^{-9}$ shows the directivity settling at an upper bound.

\section{Conclusion}\label{conclusion}
This paper has introduced a communication model for \acp{LIS} that incorporates mutual coupling.
It has shown the potential of super-directivity for \acp{LIS}. The system is able to realize theoretically unbounded directivity by overpopulating the \ac{LIS} with closely spaced antenna elements. The perfor\textbf{}mance is however limited by computational precision. Two types of antenna elements are considered, isotropic and planar. For high inter-element spacing, the directional antennas provide higher performance in the broadside direction. Specifically for broadside, at lower inter-element spacing, the two antenna types have nearly identical performance.
Next, we will extend those results to more realistic antenna patterns and feature other forms of surface, for example cylindrical or spherical. 

\section*{Acknowledgements}
This work has been supported by the Danish Council for Independent Research DFF-701700271.

\bibliographystyle{setup/IEEEtran}
\bibliography{setup/IEEEabrv,setup/IEEEexample}

\begin{thebibliography}{10}
\providecommand{\url}[1]{#1}
\csname url@samestyle\endcsname
\providecommand{\newblock}{\relax}
\providecommand{\bibinfo}[2]{#2}
\providecommand{\BIBentrySTDinterwordspacing}{\spaceskip=0pt\relax}
\providecommand{\BIBentryALTinterwordstretchfactor}{4}
\providecommand{\BIBentryALTinterwordspacing}{\spaceskip=\fontdimen2\font plus
\BIBentryALTinterwordstretchfactor\fontdimen3\font minus
  \fontdimen4\font\relax}
\providecommand{\BIBforeignlanguage}[2]{{%
\expandafter\ifx\csname l@#1\endcsname\relax
\typeout{** WARNING: IEEEtran.bst: No hyphenation pattern has been}%
\typeout{** loaded for the language `#1'. Using the pattern for}%
\typeout{** the default language instead.}%
\else
\language=\csname l@#1\endcsname
\fi
#2}}
\providecommand{\BIBdecl}{\relax}
\BIBdecl

\bibitem{marzetta2010}
T.~L. {Marzetta}, ``Noncooperative cellular wireless with unlimited numbers of
  base station antennas,'' \emph{IEEE Transactions on Wireless Communications},
  vol.~9, no.~11, pp. 3590--3600, November 2010.

\bibitem{carvalhoHeath}
\BIBentryALTinterwordspacing
E.~de~Carvalho, A.~Ali, A.~Amiri, M.~Angjelichinoski, and {Robert W. Heath
  Jr.}, ``Non-stationarities in extra-large scale massive {MIMO},''
  \emph{CoRR}, vol. abs/1903.03085, 2019. [Online]. Available:
  \url{http://arxiv.org/abs/1903.03085}
\BIBentrySTDinterwordspacing

\bibitem{abs-1901-01458}
\BIBentryALTinterwordspacing
N.~Shlezinger, O.~Dicker, Y.~C. Eldar, I.~Yoo, M.~F. Imani, and D.~R. Smith,
  ``Dynamic metasurface antennas for uplink massive {MIMO} systems,''
  \emph{CoRR}, vol. abs/1901.01458, 2019. [Online]. Available:
  \url{http://arxiv.org/abs/1901.01458}
\BIBentrySTDinterwordspacing

\bibitem{prather}
D.~W. {Prather}, S.~{Shi}, G.~J. {Schneider}, P.~{Yao}, C.~{Schuetz},
  J.~{Murakowski}, J.~C. {Deroba}, F.~{Wang}, M.~R. {Konkol}, and D.~D. {Ross},
  ``Optically upconverted, spatially coherent phased-array-antenna feed
  networks for beam-space mimo in 5g cellular communications,'' \emph{IEEE
  Transactions on Antennas and Propagation}, vol.~65, no.~12, pp. 6432--6443,
  Dec 2017.

\bibitem{lund}
S.~{Hu}, F.~{Rusek}, and O.~{Edfors}, ``Beyond massive mimo: The potential of
  data transmission with large intelligent surfaces,'' \emph{IEEE Transactions
  on Signal Processing}, vol.~66, no.~10, pp. 2746--2758, May 2018.

\bibitem{whatIsNext}
E.~Björnson, L.~Sanguinetti, H.~Wymeersch, J.~Hoydis, and T.~L. Marzetta,
  ``Massive mimo is a reality -- what is next? five promising research
  directions for antenna arrays,'' 2019.

\bibitem{Schelkunoff}
S.~A. {Schelkunoff}, ``A mathematical theory of linear arrays,'' \emph{The Bell
  System Technical Journal}, vol.~22, no.~1, pp. 80--107, Jan 1943.

\bibitem{balanis}
C.~A. Balanis, \emph{Antenna Theory: Analysis and Design}, 4th~ed.\hskip 1em
  plus 0.5em minus 0.4em\relax Wiley, 2016.

\bibitem{mailloux}
R.~J. Mailloux, \emph{Phased Array Antenna Handbook}, 2nd~ed., ser. Artech
  House antennas and propagation library.\hskip 1em plus 0.5em minus
  0.4em\relax Artech House, 2005.

\bibitem{parsons}
J.~D. Parsons, \emph{The Mobile Radio Propagation Channel}, 2nd~ed.\hskip 1em
  plus 0.5em minus 0.4em\relax John Wiley \& Sons, Ltd, 2000.

\bibitem{griffiths}
D.~J. Griffiths, \emph{Introduction to Electrodynamics}, 4th~ed.\hskip 1em plus
  0.5em minus 0.4em\relax Prentice Hall, 2013.

\bibitem{Richards2014}
M.~A. Richards, \emph{Fundamentals of Radar Signal Processing}, 2nd~ed.\hskip
  1em plus 0.5em minus 0.4em\relax McGraw-Hill Education, 2014.

\bibitem{8319526}
S.~{Hu}, F.~{Rusek}, and O.~{Edfors}, ``Beyond massive mimo: The potential of
  data transmission with large intelligent surfaces,'' \emph{IEEE Transactions
  on Signal Processing}, vol.~66, no.~10, pp. 2746--2758, May 2018.

\end{thebibliography}

\end{document}